\documentclass[preprint,pre,11pt,tightenlines]{revtex4}%
\usepackage{amsfonts}
\usepackage{amsmath}
\usepackage{amssymb}
\usepackage{graphicx}%
\providecommand{\U}[1]{\protect\rule{.1in}{.1in}}

\begin{document}
\title{Wien Effect on Ionic Conductance of Binary Strong Electrolyte Solutions in a
High External Electric Field}
\author{Byung Chan Eu and Hui Xu}
\affiliation{Department of Chemistry, McGill University, 801 Sherbrooke St. West, Montreal,
QC H3A 2K6, Canada}
\author{Kyunil Rah}
\affiliation{IT and Electronic Materials R \& D, LG Chem Research Park, 104-1 Moonji-dong,
Yuseong-gu, Daejeon 305-380, Korea }


\begin{abstract}
In the preceding paper, the exact solution of Stokes equation was obtained for
a binary strong electrolyte solution in an external electric field. In the
present paper, the solution is applied to calculate the Wien effect on
deviation from the Coulombic law of conduction in high fields. One of the
important aims of the present line of work was in removing or avoiding the
divergence difficulty in calculating the electrophoretic and relaxation time
coefficients. The present work achieves that aim by calculating on the basis
of computing the axial velocity profiles the position of the center of the ion
atmosphere as a function of the reduced field strength and therewith computing
the electrophoretic and relaxation time coefficients for the migrating
spherical ion atmosphere at each value of the reduced field strength. With the
electrophoretic and relaxation time coefficients thus calculated along the
trajectory of the center of the ion atmosphere with regard to the reduced
field strength, the equivalent ionic conductance of a magnesium sulfate
solution is calculated and compared with experimental data in very good
accuracy. It is also compared with the prediction by Wilson's theory and is
found very much improved over the latter. Moreover, the present method is
divergence-free unlike Wilson's theory. It therefore not only sheds a new
light on the nonequilibrium liquid structure of ionic solutions and the motion
of ion atmosphere in an electric field, but also shows how the Onsager--Wilson
theory of conductance in electrolyte solutions should be applied to account
for the Wien effect on ionic conductance.

\end{abstract}
\date[Date text]{:
\today
}
\startpage{1}
\endpage{102}
\maketitle

\section{Introduction}%

\setlength{\baselineskip}{20pt}
In the preceding companion paper\cite{eurah1}, henceforth referred to as paper
I or simply I, we have obtained the exact solutions for the governing
equations for nonequilibrium pair distribution functions and potentials in a
Brownian motion model, and also the solutions for the Stokes
equation\cite{stokes, landau,batchelor}, which yield the velocity of the
countercurrent induced by the motion of ions in a binary strong electrolyte
solutions under the influence of an external electric field. Therewith we
obtained the general formulas for the electrophoretic and relaxation time
factors that can be made use of to define the divergence-free electrophoretic
and relaxation time coefficients with which we may calculate ionic conductance
of binary electrolyte solutions. Since the electrophoretic and relaxation time
factors, $\mathfrak{f}(x,r;\xi)$ and $\mathfrak{g}(x,r;\xi)$, obtained are for
all allowable values of position ($x$ and $r$), it is necessary to select an
appropriate position or a set of positions that must yield divergence-free
electrophoretic and relaxation time coefficients. Since $x$ and $r$ represent
the position of the center ion of the ion atmosphere at the given value of
external field strength $\xi$ we have first determined the shape of ion
atmosphere from the projection $C_{P}$ of the axial velocity profile and
obtained the coordinates of its center position. Then therewith we can
calculate the electrohoretic and relaxation time coefficients as functions of
external field strength $\xi$. The salient point of the procedure put forward
here is that since the center position ($x_{c},0$) of the ion atmosphere moves
along the $x$ axis from the origin as $\xi$ increases, the electrophoretic
coefficient for the center ion must be calculated for the moving spherical ion
atmosphere at each value of $\xi$, which follows the trajectory described in
Fig. 7 of paper I, and therewith the ionic conductance. In this connection, it
should be recalled that Wilson\cite{wilson} evaluated the electrophoretic
coefficient\ for the center ion located at the coordinate origin $x=r=0$ for
all values of $\xi$, which consequently gives rise to a divergent term that
must be discarded to obtain a finite electrophoretic coefficient. Such a
divergent integral arises mainly because $x=r=0$ is taken for the center ion
position for the ion atmosphere even if $\xi>0$. Moreover, his procedure
ignores the fact that the ion atmosphere not only is distorted, but also
migrates in the field direction under the influece of an external field.
Therefore the coordinate position in the electrophoretic and relaxation time
factors must be other than $x=r=0$ and also a function of $\xi$.

In the present paper, we would like to apply this important point of
observation to calculate the electrophoretic coefficient and the corresponding
relaxation time coefficient, and therewith the equivalent ionic conductance in
the nonlinear regime of the field strength as a function of $\xi$; in
particular, the Wien effect\cite{wien} on equivalent conductance. Thus we
would like to show that the Onsager--Wilson
theory\cite{onsager,onsager3,wilson} is basically capable of predicting in an
excellent accuracy a marked deviation from Coulomb's law of conduction if we
avoid Wilson's procedure of taking $x=r=0$ in the electrophoretic and
relaxation time coeffficient formulas. When they are calculated according to
our procedure, the equivalent ionic conductance compares well with
experimental data on magnesium sulfate solution available in the
literature\cite{eckstrom}. As a matter of fact, when the electrophoretic
coefficient $\mathfrak{f}\left(  \xi\right)  $ and the relaxation time
coefficient $\mathfrak{g}\left(  \xi\right)  $ thus calculated according to
our procedure, the present method becomes not only divergence-free, but also
improves the accuracy of the Onsager--Wilson theory qualitatively and
quantitatively over Wilson's original theoretical prediction made with $x=r=0$
used for the center position for all values of $\xi$.

The present paper is organized as follows: In Sec. II, the electrophoretic and
the relaxation time factors are briefly reviewed and the corresponding
coefficients are defined. In Sec. III, the constitutive equations for
velocities in terms of diffusion coefficients and mean external forces are
discussed within the framework of thermodynamics of linear irreversible
processes. In Sec. IV, the theory of ionic conductance is presented and, upon
use of the electrophoretic and relaxation time coefficients defined according
to the procedure mentioned, the relative ionic conductance is calculated as a
function of reduced external field strength $\xi$ for magnesium sulfate
solution in excellent agreement with experimental data. Sec. V is for
discussion and conclusion.

\section{Electrophoretic and Relaxation Time Factors}

\subsection{Electrophoretic Coefficient}

From the axial velocity $\mathbf{v}_{x}\left(  x,r;\xi\right)  $ completely
calculated from the formal Fourier transform solutions we have shown the
electrophoretic factor $\mathfrak{f}\left(  x,r;\xi\right)  $ may be
identified:%
\begin{equation}
\mathbf{v}_{x}\left(  x,r;\xi\right)  =-\frac{k_{B}T\kappa^{2}}{8\pi\eta_{0}%
}\frac{x}{\left(  x^{2}+r^{2}\right)  ^{3/2}}-\frac{zeX\kappa}{6\sqrt{2}%
\pi\eta_{0}}\mathfrak{f}\left(  x,r;\xi\right)  , \label{vx1}%
\end{equation}
where $\mathfrak{f}\left(  x,r;\xi\right)  $ is given by%
\begin{align}
\mathfrak{f}(x,r;\xi)  &  =\frac{3}{2\left(  x^{2}+r^{2}\right)  ^{1/2}}%
-\frac{3r^{2}}{4\left(  r^{2}+x^{2}\right)  ^{3/2}}-\frac{3\left(
2x^{2}-r^{2}\right)  }{2\left(  r^{2}+x^{2}\right)  ^{5/2}}\nonumber\\
&  +\frac{3}{2}\left[  \mathfrak{C}_{1}\left(  x,r,\xi\right)  +2\mathfrak{C}%
_{2}\left(  x,r,\xi\right)  -\mathfrak{S}_{1}\left(  x,r,\xi\right)
-2\mathfrak{S}_{2}\left(  x,r,\xi\right)  \right]  \label{fephc}%
\end{align}
with $\mathfrak{C}_{1}\left(  x,r,\xi\right)  $, $\mathfrak{C}_{2}\left(
x,r,\xi\right)  $, $\mathfrak{S}_{1}\left(  x,r,\xi\right)  $, and
$\mathfrak{S}_{2}\left(  x,r,\xi\right)  $ defined by the integrals%
\begin{align}
\mathfrak{C}_{1}\left(  x,r;\xi\right)   &  =\int_{0}^{\sqrt{2\left(
1+\xi^{2}\right)  }}dy\frac{\left(  1-y^{2}+\sqrt{1+2\xi^{2}y^{2}}\right)
}{1+2\xi^{2}y^{2}}e^{-xy}I_{0}(\overline{\omega}_{1}r),\label{vC1}\\
\mathfrak{C}_{2}\left(  x,r;\xi\right)   &  =\int_{0}^{1}dy\frac{2\xi^{2}%
y^{2}\left(  1-y^{2}\right)  }{1+2\xi^{2}y^{2}}e^{-xy}I_{0}(\overline{\omega
}_{3}r),\label{Vc2}\\
\mathfrak{S}_{1}\left(  x,r;\xi\right)   &  =\int_{0}^{\sqrt{2\left(
1+\xi^{2}\right)  }}dy\frac{\sqrt{2}\xi y\left(  1-y^{2}+\sqrt{1+2\xi^{2}%
y^{2}}\right)  }{\left(  1+2\xi^{2}y^{2}\right)  \left(  1+\sqrt{1+2\xi
^{2}y^{2}}\right)  }e^{-xy}I_{0}(\overline{\omega}_{1}r),\label{vC3}\\
\mathfrak{S}_{2}\left(  x,r;\xi\right)   &  =\int_{0}^{1}dy\frac{\sqrt{2}\xi
y\left(  1-y^{2}\right)  }{1+2\xi^{2}y^{2}}e^{-xy}I_{0}(\overline{\omega}%
_{3}r). \label{vC4}%
\end{align}
It is useful to remind the reader that the second term on the right of Eq.
(\ref{vx1}) is in the same form as the electrophoretic effect term---i.e., the
last term in Eq. (3) of paper I---in the velocity formula obtained on the
basis of a heuristic argument. Here the notation is the same as in Paper I,
but we reiterate that $\xi=zeX/\kappa k_{B}T$ is the reduced external field
strength and
\begin{align}
\overline{\omega}_{1}  &  =\sqrt{1-y^{2}+\sqrt{1+2\xi^{2}y^{2}}}%
,\quad\overline{\omega}_{2}=\sqrt{1-y^{2}-\sqrt{1+2\xi^{2}y^{2}}},\nonumber\\
\overline{\omega}_{3}  &  =\sqrt{1-y^{2}},\quad\overline{\omega}=\frac
{\sqrt{1+2\xi^{2}}}{\sqrt{2}\xi}. \label{omeg}%
\end{align}
Therefore, we see that $\mathfrak{f}\left(  x,r;\xi\right)  $ is a measure of
variation\cite{mazurEPC} in the external body-force on the nonequilibrium ion
atmosphere that has resulted from the originally spherical ion atmosphere of
radius $\left(  1/\sqrt{2}\right)  \kappa^{-1}$ (in reduced units).

In Fig. 1, an example of a three-dimensional surface of $\mathfrak{f}\left(
x,r;\xi\right)  $ in the case of $\xi=3$ is plotted and the projections of the
level curves on the surface are plotted in Fig. 2. The level curves appear as
quasi-ellipses in the ($x,r$) plane. In particular, the outermost curve
$C_{P}$ denotes the locus of zero of the surface as the intersection with the
($x,r$) plane. The appearance of two quasi-ellipses making up the locus
$C_{P}$ implies that there are two velocity components, axial and transversal,
developing as the external field is increased from zero. It also means that
the spherical ion atmosphere formed in the electrolyte solution at equilibrium
when the external field is absent deforms into two quasi-ellipses, one along
the $x$ axis and the other along the $r$ axis transversal to the $x$ axis.
(This phenomenon in fact gives rise to a Hall effect. It is interesting, but
will not be a subject of investigation in this paper. We hope to return to
this aspect in a future work. Our main interest has been and is the
quasi-ellipsoid along the axial field direction and its associated ion
atmosphere in the present series of work.) As is evident from Fig. 1 and Fig.
2, the spherical ion atmosphere is now distorted into a quasi-ellipse (a
quasi-ellipsoid if rotated about the $x$ axis which is the symmetry axis)
described by the level curve $C_{p}$ with its center displaced to $\left(
x_{c},0\right)  $ and tracing the trajectory in the ($x_{c},\xi$) plane shown
in Fig. 7 of paper I and also in Table 1, as $\xi$ increases.
\begin{table}[tbp] \centering
\caption{Trajectory of the center position of the quasi-ellipse\newline
as the reduced field strength increases.}%
\begin{tabular}
[c]{|c|c|c|c|}\hline
$x_{c}$ & $\xi$ & $x_{c}$ & $\xi$\\\hline\hline
0.2 & 0.5670 & 2.4 & 0.5960\\
0.4 & 0.5795 & 2.6 & 0.5930\\
0.6 & 0.5900 & 2.8 & 0.5910\\
0.8 & 0.5975 & 3.0 & 0.5895\\
1.0 & 0.6050 & 3.2 & 0.5890\\
1.2 & 0.6070 & 3.4 & 0.5885\\
1.4 & 0.6080 & 3.6 & 0.5880\\
1.6 & 0.6075 & 3.8 & 0.5880\\
1.8 & 0.6065 & 4.0 & 0.5875\\
2.0 & 0.6030 & 4.2 & 0.5875\\
2.2 & 0.5995 & 4.4 & 0.5870\\\hline\hline
\end{tabular}
\label{Table 1}%
\end{table}
Since it would be useful to discuss the feature of $\mathfrak{f}(x,r;\xi)$ in
a little more detail, we would like to elaborate on it below:

Since the exact velocity profile consists of a purely mechanical part $\left(
\widehat{\mathbf{v}}_{x}\right)  _{\text{me}}$ and the Brownian motion (i.e.,
stochastic) part $\left(  \widehat{\mathbf{v}}_{x}\right)  _{\text{Brown}}$,
which are opposite in their numerical effects in some region of space, the
electrophoretic factor $\mathfrak{f}(x,r;\xi)$, which is a surface in the
plane of coordinates $x$ and $r$, has a locus of zero defined by
$\mathfrak{f}(x,r;\xi)=0$; see curve $C_{P}$ in Fig. 2 of this paper and also
Fig. 6 of paper I. The center of the region bounded by this locus $C_{P}$ not
only migrates along the $x$ axis (the field direction) as the field strength
increases, but also $C_{P}$ changes its shape as $\xi$ varies. When
graphically examined, the coordinate $x_{c}$ of the center of $C_{P}$ traces a
certain trajectory as $\xi$ increases from $\xi=0$, as shown in Fig. 7 of
paper I. Furthermore, $\mathfrak{f}(x,r;\xi)<0$ if $\left(  x,r\right)  $ is
within the domain defined by $C_{P}$ whereas $\mathfrak{f}(x,r;\xi)>0$ if
$\left(  x,r\right)  $ is exterior to $C_{P}$. This feature implies that the
spherically symmetric ion atmosphere at $\xi=0$ (i.e., equilibrium) with the
center at the coordinate origin polarizes to a negative velocity domain within
$C_{P}$ and a positive velocity domain exterior to $C_{P}$ in the velocity
space while the center of the quasi-elliptic domain of $C_{P}$ at $\left(
x_{c},0\right)  $, where $x_{c}>0$, is displaced from the coordinate origin
where the center of equilibrium ion atmosphere was located. Therefore, the
spherical ion atmosphere displays two regions of opposite signs and the
conductance of ions is contributed by ions in the exterior shell, and the
electrophoretic coefficient should be calculated with the force by the flowing
medium acting on the sphere (i.e., ion atmosphere) of radius $x_{c}$, part of
which is exterior to $C_{P}$, but centered at $(x_{c},0)$. This means the
transversal distance to the $x$ axis (i.e., the radial distance) on the
spherical ion atmosphere of radius $x_{c}$ (in reduced units used here) in
question is $r=x_{c}$ from the coordinate origin (the original center).
Therefore, the electrophoretic coefficient at $\xi>0$ for the spherical ion
atmosphere at $(x_{c},0)$ is given by $\mathfrak{f}(\xi)\equiv\mathfrak{f}%
(x_{c},r_{c};\xi)=\mathfrak{f}(x_{c},x_{c};\xi)$, if the Stokes
law\cite{stokes, landau,batchelor} is applied to this spherical ion
atmosphere. This, in essence, is the content of the aforementioned procedure
for selecting $x$ and $r$ in the electrophoretic factor $\mathfrak{f}%
(x,r;\xi)$. This procedure makes the selected values for $x=x_{c}$ and
$r=x_{c}$ unique because there is only one center and one spherical ion
atmosphere of radius $(1/\sqrt{2})\kappa^{-1}$ for each value of $\xi$. Note,
of course, the set $(x_{c},r_{c})$ is axially symmetric, because the system is
axially symmetric around the $x$ axis, which is parallel to the field direction.

Therefore, the Stokes law should be applied to the spherical ion atmosphere of
radius $x_{c}$ in reduced units, whose center is located at $\left(
x_{c},0\right)  $. Thus, the electrophoretic coefficient at $\xi$ is defined
by%
\begin{equation}
\mathfrak{f}\left(  \xi\right)  \equiv\mathfrak{f}\left(  x_{c},r_{c}%
;\xi\right)  =\mathfrak{f}\left(  x_{c},x_{c};\xi\right)  . \label{epf}%
\end{equation}
It should be noted that $\mathfrak{f}\left(  x,r;\xi\right)  $ now is
everywhere finite for $x>0$ and $r>0\ $for all $\xi>0$. We have calculated
$\mathfrak{f}\left(  \xi\right)  $ defined in Eq. (\ref{epf}) according to the
procedure formulated in paper I and mentioned earlier. In Table 1, the
dependence on $\xi$ of $x_{c}$ in the reduced units employed in this work is
tabulated for the purpose of future use in computing the $\mathfrak{f}\left(
\xi\right)  $ and $\mathfrak{g}\left(  \xi\right)  $ for binary electrolytes.
It is interesting to note that this trajectory has a maximum and thereafter
diminishes to an asymptote as the field strength increases.

\subsection{Relaxation Time Coefficient}

In paper I, it is shown that the local ionic field $e_{j}\Delta\mathbf{X}$ can
be obtained from the solutions of the governing equations in the form%
\begin{equation}
e_{j}\Delta X\left(  \mathbf{r}\right)  =-\frac{e_{j}^{2}\kappa^{2}}{2D}%
g_{s}(x,r;0)-\frac{e_{j}\xi\kappa^{2}}{2D}\mathfrak{g}\left(  x,r;\xi\right)
, \label{IF}%
\end{equation}
where the relaxation time factor $\mathfrak{g}\left(  x,r;\xi\right)  $ is
given by the formula%
\begin{equation}
\mathfrak{g}\left(  x,r;\xi\right)  =g_{c}\left(  x,r;\xi\right)  +\Delta
g_{s}(x,r;\xi) \label{g}%
\end{equation}
with $g_{s}(x,r;0)$, $g_{c}\left(  x,r;\xi\right)  $ and $\Delta g_{s}%
(x,r;\xi)$, respectively, defined by the integrals%
\begin{align}
g_{s}(x,r;0)  &  =\int_{0}^{\sqrt{2}}dye^{-xy}yI_{0}\left(  r\sqrt{2-y^{2}%
}\right)  ,\label{gs0}\\
g_{c}\left(  x,r;\xi\right)   &  =\frac{1}{\sqrt{2}}\int_{0}^{\sqrt{2\left(
1+\xi^{2}\right)  }}dy\frac{e^{-xy}y^{2}I_{0}\left(  \overline{\omega}%
_{1}r\right)  }{1+2\xi^{2}y^{2}}-\sqrt{2}\int_{0}^{1}dy\frac{e^{-xy}y^{2}%
I_{0}\left(  \overline{\omega}_{3}r\right)  }{1+2\xi^{2}y^{2}},\label{gc}\\
\Delta g_{s}(x,r;\xi)  &  =\frac{1}{\xi}\left[  \frac{1}{2}\int_{0}%
^{\sqrt{2\left(  1+\xi^{2}\right)  }}dy\frac{e^{-xy}y\left(  1+\sqrt
{1+2\xi^{2}y^{2}}\right)  I_{0}\left(  \overline{\omega}_{1}r\right)  }%
{1+2\xi^{2}y^{2}}\right. \nonumber\\
&  \left.  -\int_{0}^{\sqrt{2}}dye^{-xy}yI_{0}\left(  r\sqrt{2-y^{2}}\right)
\right]  +2\xi\int_{0}^{1}dy\frac{y^{3}e^{-xy}I_{0}\left(  \overline{\omega
}_{3}r\right)  }{1+2\xi^{2}y^{2}}. \label{gs}%
\end{align}
It should be noted that $\Delta g_{s}(x,r;\xi)$ tends to a constant as
$\xi\rightarrow0$. The integrals in Eqs. (\ref{gc}) and (\ref{gs}) are also
subject to Ineq. (90) or (A20) of paper I. Using this form of relaxation time
factor $\mathfrak{g}\left(  x,r;\xi\right)  $, we have shown in Paper I that
Wilson's formula\cite{wilson,harned} $g(\xi)$ for the relaxation time
coefficient can be recovered as an approximation if both $x$ and $r$ are taken
equal to zero. It is an approximation since the contribution from $\Delta
g_{s}(x,r;\xi)$ must be discarded although its magnitude is not negligible
when the results of the integrals are evaluated at $x=r=0$.

To be consistent with the electrophoretic coefficient defined earlier, we
define the relaxation time coefficient $\mathfrak{g}\left(  \xi\right)  $ with
the relation%
\begin{equation}
\mathfrak{g}\left(  \xi\right)  =g_{c}\left(  x_{c},r_{c};\xi\right)  +\Delta
g_{s}(x_{c},r_{c};\xi)=g_{c}\left(  x_{c},x_{c};\xi\right)  +\Delta
g_{s}(x_{c},x_{c};\xi). \label{gf}%
\end{equation}
This relaxation time coefficient $\mathfrak{g}\left(  \xi\right)  $ will be
used in the calculation of ionic conductance as a function of the external
field strength and, in particular, the Wien effect on the ionic
conductance\cite{eckstrom,harned} of magnesium sulfate solution. In Fig. 3 and
Fig. 4, $\mathfrak{f}\left(  \xi\right)  $ and $\mathfrak{g}\left(
\xi\right)  $ defined in Eqs. (\ref{epf}) and (\ref{gf}) are, respectively,
calculated and compared with the Wilson's results. Compare these results for
$\mathfrak{f}\left(  \xi\right)  $ and $\mathfrak{g}\left(  \xi\right)  $ with
the examples for $\mathfrak{f}\left(  x,r;\xi\right)  $ and $\mathfrak{g}%
\left(  x,r;\xi\right)  $ for $x=r=0.5$ shown in Figs. 8 and 10, respectively,
of Paper I. The values of $x$ and $r$ chosen are arbitrary.

\section{Constitutive Equation for Diffusion and Conductance}

Macroscopic description of ionic motions in solution in an external electric
field must be subjected to the laws of irreversible thermodynamics. Since the
macroscopic process of interest here is steady, we first assume that it obeys
steady-state linear irreversible thermodynamic constitutive
equations---namely, linear thermodynamic force--flux relations. On close
examination of the OW theory\cite{wilson} it is evident that the linear
irreversible thermodynamic constitutive equations for diffusion underlie it.
If they are found inadequate for constitutive equations in the linear regime
then we will have to seek a way to extend them to the nonlinear regime. For
this, see, for example, Refs. 14 and 15.

Here the ionic diffusion flux $\mathbf{J}_{k}$ of ion $k$ is assumed to follow
a linear constitutive relation according to the linear irreversible
thermodynamics\cite{mazur,haase}. According to the kinetic theory of
fluids\cite{eu87,eubk2,chapman}, for binary mixtures at steady state the
thermodynamic force $\mathbf{d}_{k}$ for diffusion is given by the linear
thermodynamic force%
\begin{equation}
\mathbf{d}_{k}=x_{k}\mathbf{\nabla}p_{k}-\frac{\rho_{k}}{p}\mathbf{F}_{k},
\label{30}%
\end{equation}
where $x_{k}=n_{k}/n$ is the mole fraction, $p_{k}$ is the partial pressure,
$p$ is the pressure, $\rho_{k}$ is the mass density of species $k$, and
$\mathbf{F}_{k}$ is the external force on the species per mass. In the
conductance experiment considered in this work, the partial pressure is
uniform in space, so that the pressure gradient is equal to zero in first
order. In the case of a binary electrolyte solution in which the continuum
solvent is assumed spatially uniform, there is only one independent diffusion
coefficient, and the linear thermodynamic force--flux relation or the
constitutive equation for diffusion may be written as%
\begin{equation}
\mathbf{J}_{k}=-\omega_{k}p\mathbf{d}_{k}, \label{31}%
\end{equation}
where $\omega_{k}$ is the diffusion coefficient of species $k$. Since the
diffusion flux is defined relative to a reference velocity by the
formula\cite{mazur,chapman}%
\begin{equation}
\mathbf{J}_{k}=\mathbf{v}_{k}-\mathbf{u,} \label{32}%
\end{equation}
where $\mathbf{u}$ is the reference velocity suitably chosen, such as the
barycentric velocity, which obeys the hydrodynamic equations of the solution.
There are other possibilities chosen for the reference velocity in the
literature\cite{mazur}. However, in the theory of conductance in electrolyte
solutions the barycentric velocity is generally used, and if the kinetic
theory\cite{eu87,eubk2,chapman} of diffusion is developed, $\mathbf{u}$ is
exactly what appears in the equation for the diffusion flux. Since we are
generally interested in mobilities in the conductance experiments, it is
required to calculate $\mathbf{v}_{k}$ itself of species $k$ instead of the
diffusion flux $\mathbf{J}_{k}$. Thus we have%
\begin{equation}
\mathbf{v}_{k}=-\omega_{k}p\mathbf{d}_{k}+\mathbf{v}\left(  \mathbf{r}\right)
=\omega_{k}\rho_{k}\mathbf{F}_{k}+\mathbf{v}\left(  \mathbf{r}\right)  ,
\label{33}%
\end{equation}
in which we have taken $\mathbf{u\equiv v}\left(  \mathbf{r}\right)  $, the
hydrodynamic velocity calculated previously\cite{eurah1} and presented in the
previous section. It should be noted that there is no arbitrariness about the
identification of $\mathbf{u}$ with $\mathbf{v}\left(  \mathbf{r}\right)  $,
because generally $\mathbf{u}$ in Eq. (\ref{32}) is certainly the barycentric
velocity obeying the momentum balance equation---the Navier--Stokes
equation\cite{landau} or the Stokes equation\cite{stokes,landau,batchelor} at
the level of linear irreversible thermodynamics. By this consideration, we now
see that the electrophoretic coefficient considered in the theory of
conductance is intimately associated with the hydrodynamic flow velocity
$\mathbf{v}\left(  \mathbf{r}\right)  $, which is the solution of the Stokes
equation for flow of the medium in the binary electrolyte solution in the
present theory. However, this hydrodynamic velocity to be used is that
evaluated at $\left(  x_{c},r_{c}\right)  $ at which the electrophoretic
coefficient $\mathfrak{f}\left(  x_{c},r_{c};\xi\right)  $ is computed.

\section{Conductance}

The velocity $\mathbf{v}_{k}$ of ion $k$ at $\left(  x_{c},r_{c}\right)  $ is
computed relative to a suitable reference velocity---for example, the solution
of the Stokes equation. Identifying $\rho_{k}\mathbf{F}_{k}$ with
$e_{k}\mathbf{X}_{t}\mathbf{=\,}e_{k}\mathbf{X+}e_{k}\Delta\mathbf{X}$ and
taking the reference velocity $\mathbf{u}\left(  x_{c},r_{c}\right)  $ with
the axial velocity $\mathbf{v}_{x}\ $at $\left(  x_{c},r_{c}\right)  $, we
find%
\begin{equation}
\mathbf{v}_{k}\left(  \xi\right)  =\omega_{k}e_{k}X-\frac{\omega_{k}e_{k}%
\mu^{\prime}\kappa}{2D}\mathfrak{g}\left(  \xi\right)  -\frac{\left\vert
z_{k}\right\vert eX\kappa}{6\sqrt{2}\pi\eta_{0}}\mathfrak{f}\left(
\xi\right)  \quad\left(  k=i,j\right)  . \label{34}%
\end{equation}
Here we have taken the ionic field $\Delta\mathbf{X}$ as for the relaxation
time effect in Eq. (\ref{IF})%
\begin{equation}
e_{j}\Delta X=-\frac{e_{j}\xi\kappa^{2}}{2D}\mathfrak{g}\left(  x_{c}%
,r_{c};\xi\right)  =-\frac{e_{j}\xi\kappa^{2}}{2D}\mathfrak{g}\left(
\xi\right)  \label{34g}%
\end{equation}
without the field-independent term which has nothing to do with the relaxation
of distorted ion atmosphere, because the relaxation time coefficient is
representative of the relaxation of asymmetry of the ionic atmosphere.
Similarly, the electrophoretic effect is represented by%
\begin{equation}
\mathbf{v}_{x}\left(  \xi\right)  =-\frac{zeX\kappa}{6\sqrt{2}\pi\eta_{0}%
}\mathfrak{f}\left(  x_{c},r_{c};\xi\right)  =-\frac{zeX\kappa}{6\sqrt{2}%
\pi\eta_{0}}\mathfrak{f}\left(  \xi\right)  \label{34f}%
\end{equation}
without the field-independent term.

The mobility in electrostatic units is then given by\cite{harned}%
\begin{equation}
u_{k}=\frac{\mathbf{v}_{k}\left(  \xi\right)  }{X}, \label{40}%
\end{equation}
and it may be written in practical units as%
\begin{equation}
u_{k}=\frac{1}{300}\left(  \left\vert e_{k}\right\vert \omega_{k}-\frac
{\omega_{k}e_{k}\mu^{\prime}\kappa}{2D}\mathfrak{g}(\xi)-\frac{\left\vert
e_{k}\right\vert \kappa}{6\sqrt{2}\pi\eta_{0}}\mathfrak{f}(\xi)\right)  .
\label{41}%
\end{equation}
At infinite dilution, $\kappa\rightarrow0$ and the limiting mobility is given
by%
\begin{equation}
u_{k}^{0}=\frac{\left\vert e_{k}\right\vert \omega_{k}}{300}. \label{42}%
\end{equation}
Since the limiting equivalent conductance is%
\begin{equation}
\Lambda_{k}^{0}=Fu_{k}^{0}, \label{43}%
\end{equation}
where $F$ is the Faraday constant ($F=96493.1$), we have the equivalent
conductance of ion $k$ in practical units given in the form%
\begin{equation}
\Lambda_{k}=\Lambda_{k}^{0}-\frac{\left(  ze\right)  ^{2}\kappa\Lambda_{k}%
^{0}}{2Dk_{B}T}\mathfrak{g}(\xi)-\frac{F\left\vert e_{k}\right\vert \kappa
}{6\times300\sqrt{2}\pi\eta_{0}}\mathfrak{f}(\xi). \label{44}%
\end{equation}

Define the equivalent conductance of the binary electrolyte%
\begin{equation}
\Lambda=\Lambda_{+}+\Lambda_{-}\equiv\Lambda_{1}+\Lambda_{2}. \label{45a}%
\end{equation}
Then the equivalent conductance for a binary electrolyte is given by%
\begin{equation}
\Lambda\left(  \xi\right)  =\Lambda^{0}-\frac{\left(  ze\right)  ^{2}%
\kappa\Lambda^{0}}{2Dk_{B}T}\mathfrak{g}(\xi)-\frac{F\left(  \left\vert
e_{1}\right\vert +\left\vert e_{2}\right\vert \right)  \kappa}{6\times
300\sqrt{2}\pi\eta_{0}}\mathfrak{f}(\xi) \label{45}%
\end{equation}
with $\Lambda^{0}=\Lambda_{1}^{0}+\Lambda_{2}^{0}$.

In experiment, the relative equivalent conductance
\begin{equation}
\Delta\Lambda\left(  \xi\right)  =\Lambda\left(  \xi\right)  -\Lambda\left(
0\right)  \label{46}%
\end{equation}
is reported as a function of the field strength $\xi$. Since in the limit of
$\xi=0$%
\begin{equation}
\Lambda\left(  0\right)  =\Lambda^{0}-\frac{\left(  ze\right)  ^{2}%
\kappa\Lambda^{0}}{2Dk_{B}T}\mathfrak{g}(0)-\frac{F\left(  \left\vert
e_{1}\right\vert +\left\vert e_{2}\right\vert \right)  \kappa}{6\times
300\sqrt{2}\pi\eta_{0}}\mathfrak{f}(0), \label{47}%
\end{equation}
we obtain the relative equivalent conductance
\begin{equation}
\Delta\Lambda\left(  \xi\right)  =\frac{\left(  ze\right)  ^{2}\kappa
\Lambda^{0}}{2Dk_{B}T}\left[  \mathfrak{g}(0)-\mathfrak{g}(\xi)\right]
+\frac{F\left(  \left\vert e_{1}\right\vert +\left\vert e_{2}\right\vert
\right)  \kappa}{6\times300\sqrt{2}\pi\eta_{0}}\left[  \mathfrak{f}%
(0)-\mathfrak{f}(\xi)\right]  . \label{48}%
\end{equation}

This formula is computed for various values of reduced field strength $\xi$,
and $\Delta\Lambda\left(  \xi\right)  /\Lambda\left(  0\right)  $ calculated
is plotted against $\xi$ and compared with experimental
data\cite{eckstrom,harned} on a MgSO$_{4}$ solution in Fig. 5. It is the only
data of this kind available in the literature for strong binary electrolytes
which we can use for comparison on the Wien effect. We have taken
$d=2R_{\text{MgSO}_{4}}=2\times3.12=6.24$ for the diameter of MgSO$_{4}$,
$T=291$ K, $D=81$, $c=1.22\times10^{-3}$ mole/liter, $\eta_{0}=0.0105$ poise,
the limiting conductance values\cite{harned} $\lambda_{0}^{+}=53.06$ for
Mg$^{+2}$ and $\lambda_{0}^{-}=80.0$ for SO$_{4}^{-2}$, respectively, and
$120.36$ g/mole for the molecular weight for MgSO$_{4}$. In Fig. 5, the solid
curve is the prediction by the present theory, the solid squares are
experimental data, and the broken curve is the prediction by Wilson's method,
that is, the prediction with the electrophoretic coefficient $f(\xi)$ and the
relaxation time coefficient $g(\xi)$ calculated by Wilson in his
dissertation\cite{wilson,harned}. Note that they are tabulated in the
monograph of Harned and Owen\cite{harned}. The OW theory prediction is much
smaller than the experimental conductance. Furthermore, the low field ($\xi
<1$) behavior predicted by the Wilson's result appears to be markedly
different from the experimental data and the present theoretical prediction,
the latter being almost linear with respect to $\xi$, while Wilson's
prediction appears to be nonlinear, i.e., $\xi^{\alpha}$ with $\alpha>1$. We
believe that the procedure used for selecting the $x$ and $r$ used for
calculation of the electrophoretic and relaxation time coefficients and thus
the ionic conductance appears to be well supported by the comparison made.

\section{Discussion and Concluding Remarks}

Ionic liquids are ubiquitous and it is important to understand their physical
behaviors in physical, chemical, and biological systems under the influence of
external electromagnetic fields. Onsager's theory\cite{onsager} of conductance
is an important benchmark in the development of the theory of transport
processes in ionic solutions, although his theory appears to have been almost
forgotten about and paid little attention these days except one recent
work\cite{daily}. We believe it is unwarranted to ignore his basic ideas on
the subject matter because the theory seems to provide much\ valuable insight
for further development of theories of transport processes in ionic solutions
and plasmas. In the present series of papers, we have revisited and examined
the details of his theories and have been able to obtain full evaluations of
the formal results of the OW theory. The preceding\cite{eurah1} and present
papers describe our understanding of the inner workings of his theories and
their application to an experimental system. We have found that the solutions
of the OF equations\cite{fuoss}, the Poisson equations\cite{jackson}, and
Stokes equation\cite{landau} for the velocity of the medium would require a
more mathematically complete treatment than originally implemented in Wilson's
dissertation\cite{wilson}, especially, to avoid divergence difficulty
that\ was unfortunately not dealt with squarely. In paper I\cite{eurah1} we
have exactly evaluated the formulas for the axial and transversal velocities
in a binary electrolyte solution, which may be deemed one of the goals of the
theory of ionic conductance. In the present paper, we have applied the axial
velocity formula to calculate the equivalent conductance of a binary strong
electrolyte, having elucidated the appropriate set of $(x,r)$ to use by a
computation-based procedure to select a unique set of ($x,r$) as a function of
$\xi$. Although this procedure lacks a precise theoretical foundation based on
irreversible thermodynamics or hydrodynamics, it does provide us with an
unambiguous procedure to obtain correct estimates of conductance over the
entire range of $\xi$ experimentally studied. In this sense, we may conclude
that it is a practical and useful procedure for understanding the experimental
data. In any case, the procedure is founded on the observation of how the ion
atmosphere behaves in the applied external field. The study of its behavior
with respect to the field not only provides much insight into the interplay
between the long-range Coulombic interactions between charged particles and
the interaction of the ion atmosphere with the external field producing a
rather intricate ionic liquid structure, but also shows how charged species
diffuse in such a liquid under the influence of the applied electric field.
This picture of the ionic liquid behavior that we have gained in the present
paper owes to the fact that the Brownian movement model, namely, the solutions
of the governing equations, provides us with exact nonequilibrium distribution
functions, ionic potentials, and velocity profiles that can be evaluated
rigorously either numerically of analytically. This latter feature is an
indispensable aspect of the OW theory of ionic solutions we have shown
possible in this work. We emphasize that the present work disposes of the
divergence difficulty inherent to Wilson's procedure\cite{wilson}, which arose
from an inadequate treatment of the Fourier transform integrals for the
solutions, and the theory of Wien effect is thereby made practically amenable
to theoretical investigation and satisfactorily accountable theoretically.

The theoretical significance of the Onsager theory of electrolytic
conductance\cite{onsager} lies in its ability to take into account the
non-linear field dependence in the form of local body-force that is obtained
through the Onsager--Fuoss equations\cite{fuoss} for the pair distribution
functions and the Poisson equations\cite{jackson} which are coupled together.
The local body-force thus obtained is then made use of in the Stokes equation
for the velocity of the medium subjected to the external electric field. The
present paper enables us to conclude that the Onsager theory is now made
capable of treating conduction phenomena of dilute electrolyte solutions
irrespective of the applied field strength and predicts a correct behavior
qualitatively and quantitatively in contrast to the conclusion drawn in the
literature in the past that it predicts too small values for equivalent
conductance\cite{harned}. Thus it, in fact, appears to provide a great deal of
insights into diverse electrical conduction phenomena in subjects other than
traditional electrolyte solutions studied in physical chemistry, which are
lately studied under different guises and subject names, such as
semiconductors\cite{ferry,ting}, ionic liquids\cite{barthel,note2}, ion
channels\cite{ionchannel}, plasmas\cite{mason}, etc., but still involve
charged species subjected to an external electric field gradient, often rather
strong because of small system sizes commonly studied in recent years.
Therefore the basic physics underlying the phenomena in the aforementioned
subject fields is unchanged other than sizes of the systems and conditions,
and in this respect Onsager's theory can give us a valuable lesson on how to
treat ionic solutions and ionized gases in external electromagnetic fields.

In this work and also in the OW theory\cite{wilson}, the treatment of the
thermodynamic force--flux relation is still that of linear irreversible
processes, that is, the constitutive relations for ionic motion and flow of
the medium are those of the theory of linear irreversible
processes\cite{mazur}. In addition to this restriction, it is a theory limited
to sufficiently dilute electrolyte solutions. Therefore these two important
restrictions must be removed for a more comprehensive theory of electric
conductivity in experimental systems of current interest. We leave these
aspects to future studies, but just refer to reader to Refs. 14 and 15 for
nonlinear constitutive equations.\bigskip

\textbf{Acknowledgement}

This work was supported in part by the Discovery grants from the Natural
Sciences and Engineering Research Council of Canada.

Figure Captions

Fig. 1\quad The electrophoretic factor $\mathfrak{f}\left(  x,r,\xi\right)  $
is plotted in the $\left(  x,r\right)  $ plane in 3D in the case of $\xi=3$.

Fig. 2\quad The level curves of the surface $\mathfrak{f}\left(
x,r,\xi\right)  $ in Fig. 1 is projected onto the $\left(  x,r\right)  $
plane. The level curves form two sets of quasi-ellipses with the major axes on
the $x$ and $r$ axes, respectively. The outermost level curve $C_{P}$
represents the zero of $\mathfrak{f}\left(  x,r,\xi\right)  $, namely,
$\mathfrak{f}\left(  x,r,\xi\right)  =0$. This outermost level curve $C_{P}$
indicates how the spherical ion atmosphere assumed by the equilibrium ion
atmosphere when $\xi=0$ migrates from $\left(  x,r\right)  =\left(
0,0\right)  $ to a point on the positive $x$ axis and also is distorted to a
quasi-ellipse with the major axis lying on the $x$ axis by the action of the
external electric field as $\xi$ increases, and similarly to the quasi-ellipse
with the major axis lying on the $r$ axis perpendicular to the $x$ axis. The
center of the curve $C_{P}$ follows a trajectory described by the curve shown
in Fig. 7 in paper I. The trajectory grows from zero at $\xi=0$ to a maximum
and then decreases to a plateau value as $\xi$ increases. See also Table 1 of
the present paper.

Fig. 3\quad Comparison of the electrophoretic coefficients of the present and
Wilson's theories. The solid line: the present theory; the dotted line:
Wilson's result.

Fig. 4.\quad Comparison of the relaxation time coefficients of the present and
Wilson's theory. The solid line: the present theory; the dotted line: Wilson's result.

Fig. 5.\quad Relative equivalent conductance compared with experiment and the
prediction by the Onsager--Wilson formula. The solid symbols are the
experimental data on MgSO$_{4}$ solution, the solid curve the present theory,
and the broken curve the prediction by the Onsager--Wilson formula. It should
be noted that the behaviors of the relative conductance of the present and
Wilson's theories are qualitatively and also quantitatively different.\newpage%

\begin{figure}
[ptb]
\begin{center}
\includegraphics[
natheight=5.348100in,
natwidth=8.776600in,
height=2.2854in,
width=3.7332in
]%
{New 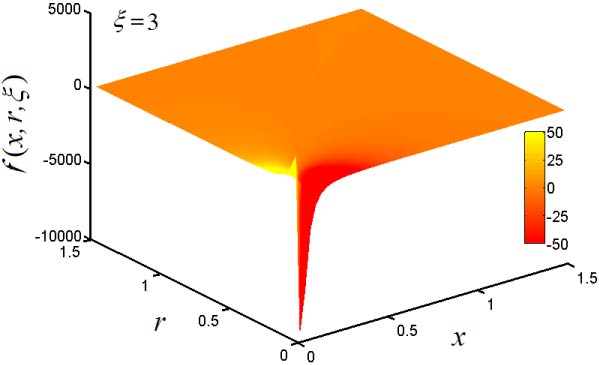}%
\caption{ }%
\label{Fig 1}%
\end{center}
\end{figure}
\newpage\
\begin{figure}
[ptb]
\begin{center}
\includegraphics[
natheight=6.725700in,
natwidth=8.527500in,
height=2.9505in,
width=3.7332in
]%
{New 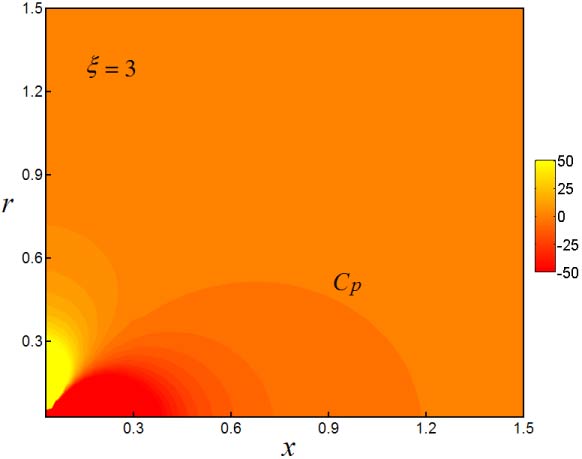}%
\caption{ }%
\label{Fig2}%
\end{center}
\end{figure}
\newpage\
\begin{figure}
[ptb]
\begin{center}
\includegraphics[
natheight=4.717600in,
natwidth=7.091500in,
height=2.4925in,
width=3.7341in
]%
{New 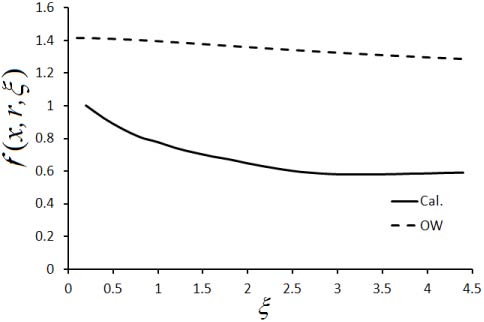}%
\caption{ }%
\label{Fig3}%
\end{center}
\end{figure}
\newpage\
\begin{figure}
[ptb]
\begin{center}
\includegraphics[
natheight=4.717600in,
natwidth=7.076900in,
height=2.4979in,
width=3.7332in
]%
{New 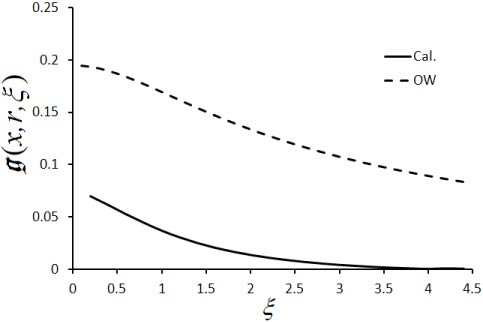}%
\caption{ }%
\label{Fig4}%
\end{center}
\end{figure}
\newpage\
\begin{figure}
[ptb]
\begin{center}
\includegraphics[
natheight=4.806100in,
natwidth=6.886200in,
height=2.6138in,
width=3.7341in
]%
{New 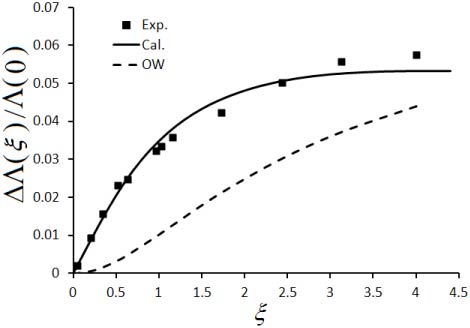}%
\caption{ }%
\label{Fig5}%
\end{center}
\end{figure}
\newpage

\end{document}